\documentclass[conference]{IEEEtran}
\IEEEoverridecommandlockouts
\usepackage{amsmath,amssymb,amsfonts}
\usepackage{algorithmic}
\usepackage{graphicx}
\usepackage{textcomp}
\usepackage{xcolor}
\usepackage{colortbl}
\usepackage{graphicx}
\usepackage{array}
\usepackage{textcomp}
\usepackage{stfloats}
\usepackage{subfigure}
\usepackage[ruled,linesnumbered]{algorithm2e}
\usepackage{authblk}
\usepackage{biblatex}
\definecolor{orange}{rgb}{0.8, 0.6, 0.2}
\definecolor{black}{rgb}{0.0, 0.0, 0.0}
\definecolor{purple}{rgb}{0.65,0,0.65}

\def\BibTeX{{\rm B\kern-.05em{\sc i\kern-.025em b}\kern-.08em
    T\kern-.1667em\lower.7ex\hbox{E}\kern-.125emX}}
\begin{document}

\title{SPSW: Database Watermarking Based on Fake Tuples and Sparse Priority Strategy
}

\author{Zhiwen Ren, Zehua Ma$^*$, Weiming Zhang$^*$, Nenghai Yu
\thanks{*Corresponding author.}\thanks{This work was supported in part by the Natural Science Foundation of China under Grant 62072421, 62002334, 62102386, 62121002 and U20B2047.}
}
\affil{School of Cyber Science and Technology \\
University of Science and Technology of China, Hefei, China\\
\{renzhiwen12@mail, mzh045@, zhangwm@, ynh@\}.ustc.edu.cn}

% \author{
% \IEEEauthorblockN{1\textsuperscript{st} Zhiwen Ren}
% \IEEEauthorblockA{\textit{School of Cyber Science and Technology} \\
% \textit{University of Science and Technology of China}\\
% Hefei, China\\
% renzhiwen12@mail.ustc.edu.cn}
% \and
% \IEEEauthorblockN{2\textsuperscript{nd} Zehua Ma}
% \IEEEauthorblockA{\textit{School of Cyber Science and Technology} \\
% \textit{University of Science and Technology of China}\\
% Hefei, China\\
% mzh045@mail.ustc.edu.cn}
% \and
% \IEEEauthorblockN{3\textsuperscript{rd} Weiming Zhang}
% \IEEEauthorblockA{\textit{School of Cyber Science and Technology} \\
% \textit{University of Science and Technology of China}\\
% Hefei, China\\
% zhangwm@ustc.edu.cn}
% \and
% \IEEEauthorblockN{4\textsuperscript{th} Nenghai Yu}
% \IEEEauthorblockA{\textit{School of Cyber Science and Technology} \\
% \textit{University of Science and Technology of China}\\
% Hefei, China\\
% ynh@ustc.edu.cn}

% \and
% \and

\maketitle

\begin{abstract}
Databases play a crucial role in storing and managing vast amounts of data in various organizations and industries. Yet the risk of database leakage poses a significant threat to data privacy and security. To trace the source of database leakage, researchers have proposed many database watermarking schemes. Among them, fake-tuples-based database watermarking shows great potential as it does not modify the original data of the database, ensuring the seamless usability of the watermarked database. However, the existing fake-tuple-based database watermarking schemes need to insert a large number of fake tuples for the embedding of each watermark bit, resulting in low watermark transparency. Therefore, we propose a novel database watermarking scheme based on fake tuples and \textbf{s}parse \textbf{p}riority \textbf{s}trategy, named SPSW, which achieves the same watermark capacity with a lower number of inserted fake tuples compared to the existing embedding strategy. Specifically, for a database about to be watermarked, we prioritize embedding the sparsest watermark sequence, i.e., the sequence containing the most `0' bits among the currently available watermark sequences. For each bit in the sparse watermark sequence, when it is set to `1', SPSW will embed the corresponding set of fake tuples into the database. Otherwise, no modifications will be made to the database. Through theoretical analysis, the proposed sparse priority strategy not only improves transparency but also enhances the robustness of the watermark. The comparative experimental results with other database watermarking schemes further validate the superior performance of the proposed SPSW, aligning with the theoretical analysis.
\end{abstract}


\section{Introduction}
\label{sec:intro}
%\subsection{Database Traceability Watermark}
Databases store sensitive and economically valuable information such as personally identifiable, financial, and customer data. The exposure of such information can lead to severe consequences, including privacy breaches, financial harm, and reputational damage  \cite{10top2022,inside2019}. However, in practical scenarios, databases are often shared among multiple users, intensifying the risk of leaks. Tracing the source of a database leak, when it occurs, is crucial for accountability and damage control. By successfully identifying the origin of the leak, responsible parties can be held accountable and appropriate measures can be taken to mitigate further harm.

As a common method of copyright protection, digital watermarking has been applied to the tracing of database leaks, and a series of solutions have been proposed \cite{gort2021semantic,li2022secure,chai2019robust,yuan2022attribute,wang2022fbipt}. Typically, to ensure traceability, the owner of the database distributes the database with different embedded watermarks to different authorized users. Then, when the database appears in an unauthorized channel, the source of the leak can be determined by extracting the embedded watermark.

The existing robust database watermarking schemes can be roughly divided into two categories: one is based on modifying the original data of the database, and the other is based on inserting fake tuples to embed the watermark. According to the type of database data, the scheme based on modification can be further divided into watermarking through character data modifications and watermarking through numerical data modifications. Schemes based on character modifications \cite{gort2021semantic,khanduja2012robust} have relatively low capacity and poor robustness, so researchers have focused more on schemes based on numerical modifications. The initial database watermark scheme based on numerical modifications was proposed by Li et al. \cite{agrawal2002watermarking}, with subsequent improvements to enhance the robustness \cite{shehab2007watermarking,cui2006robust}. Considering that watermark transparency is affected by data modification, reversible database watermarking \cite{chang2014blind,li2020robust,hwang2020reversible} and constraint optimization watermarking schemes \cite{farfoura2013novel,franco2015robust,li2019reversible,hu2018new,li2022secure,tufail2018digital} are proposed. In addition, there are optimization schemes \cite{rani2017adapting,chai2019robust,yuan2022attribute} for different scenarios, which substantially improve the watermarking performance. 

But database watermark schemes based on modifications share common limitations. They can only be implemented on databases with suitable attributes for modification, and watermarking modifications may potentially damage the original data and impact the usability of the database. Yet the second category of database watermark schemes based on the insertion of fake tuples does not have the aforementioned limitations, which have lower requirements on database attributes and avoid the modification of the original data. Fake-tuple-based database watermarking scheme\cite{ismail2021context,pournaghshband2008new,wang2022fbipt} is first proposed in \cite{pournaghshband2008new}, which can only embed a single bit and is suitable for simple application scenarios. Then, Wang et al. \cite{wang2022fbipt} proposed a novel watermark method in 2022 by inserting fake tuples under specific tuples to form specific tuple combinations to express watermark information. By grouping tuples and embedding watermarks, the scheme proposed by Wang et al. can embed multi-bit watermark information.

However, the existing schemes need to insert a considerable number of fake tuples for each embedded watermark bit, resulting in a decrease in watermark transparency. In our approach, we aim to enhance watermark transparency by optimizing the information representation and minimizing the insertion of fake tuples. Thus, in this paper, we propose a novel database watermark scheme based on \textbf{s}parse \textbf{p}riority \textbf{s}trategy called SPSW. 
The main contributions of this paper are as follows:

\begin{itemize}
    \item We propose a sparse priority strategy to better encode and embed the watermark sequence into the database with less insertion of fake tuples compared to the existing similar schemes.
    \item We quantitatively evaluated the robustness of watermarking through theoretical analysis. After theoretical derivation, we found that the proposed SPSW improves not only transparency but also robustness.
    \item Through comparative experiments with other schemes, we demonstrate the superiority of our scheme in terms of robustness and transparency.
\end{itemize}

The rest of the paper is organized as follows: Section \ref{sec:work} presents related work, then, Section\ref{sec:method} details the method, including flowcharts, pseudo-code, and textual description. Section \ref{sec:theore} theoretically analyzes the relationship between the performance and parameters. Then, Section \ref{sec:experiment} records the experimental arrangement and results, and verifies the effectiveness of the scheme. Finally, Section \ref{sec:conclusion} summarizes the proposed scheme. %full text and proposes Future suggestions, improvements, and research directions.

\section{Related Work}
\label{sec:work}
Fake-tuple-based watermark schemes hide watermark information by adding fake tuples to the database. However, the initial scheme \cite{pournaghshband2008new} can only express single-bit watermarking and cannot be applied to complex application scenarios.

To solve this problem, Wang et al. proposed an optimized fake-tuple-based scheme FBIPT \cite{wang2022fbipt}, which is capable of expressing multi-bit watermarking. This scheme uses binary strings as watermarks. When the watermark to be embedded is $L$ bits, the database tuples are correspondingly divided into $L$ groups, and each group of tuples is assigned a group number $G$, whose value ranges from 0 to $L$-1. Since the grouping rule is independent of the tuple order, neighboring tuples may have different group numbers. Then, the $j$-th bit of the watermark is embedded in the $j$-th group of tuples.

Tuples are evenly bisected into ``1-tuple" and ``0-tuple" according to certain characteristics. Two tuples are adjacent, when the ``1-tuple" is above the ``0-tuple", it is called ``10-combination"; when the ``0-tuple" is above the ``1-tuple", it is called ``01-combination". The number of ``10-combination" and ``01-combination" in each group of tuples is roughly equal, and inserting fake tuples will increase the number of certain combinations. When embedding a watermark, if the watermark bit to be embedded in a group of tuples is ``1", a ``0-tuple" will be inserted under the ``1-tuple" in the group of tuples to form an additional ``10-combination". In the same way, ``01-combination" is added to embed bit ``0" by inserting a ``1-tuple". The algorithm also adjusts the number of inserted fake tuples by controlling parameters and records the average number of inserted fake tuples in each group of tuples as $x$. The extraction algorithm determines the hidden watermark bit of each group by comparing the numbers of the two combinations. 

Table~I shows one example of a watermarked database generated by FBIPT, where some parameters have been omitted for clarity. In this example, we set $x=1$. To embed a watermark sequence whose length $L$ is 2, the algorithm divides the original tuple (black) into two groups, and the group number is marked as $G$. For every 1-bit watermark embedded, FBIPT needs to insert a fake tuple (red) to construct a combination of ``01" or ``10", and finally inserts two fake tuples. So, for the FBIPT scheme, the average number of inserted fake tuples is $xL$.

\renewcommand{\arraystretch}{1.4}

\begin{table}[]
\centering
\label{tab:FBIPT}
\caption{Embedding using the FBIPT scheme.}
\resizebox{\linewidth}{!}{
\setlength{\tabcolsep}{.1in}{
\begin{tabular}{|
>{\columncolor[HTML]{FFFFFF}}c 
>{\columncolor[HTML]{FFFFFF}}c ccc|}
\hline
\multicolumn{1}{|c|}{\cellcolor[HTML]{FFFFFF}\textbf{Filght number}}                & \multicolumn{1}{c|}{\cellcolor[HTML]{FFFFFF}\textbf{Department}}                      & \multicolumn{1}{c|}{\textbf{Day}}                              & \multicolumn{1}{c|}{\textbf{Time}}                         & \textbf{G} \\ \hline
\multicolumn{1}{|c|}{\cellcolor[HTML]{FFFFFF}Z0702}                                 & \multicolumn{1}{c|}{\cellcolor[HTML]{FFFFFF}London}                                   & \multicolumn{1}{c|}{Saturday}                                  & \multicolumn{1}{c|}{18:20}                                 & 2          \\ \hline
\multicolumn{1}{|c|}{\cellcolor[HTML]{FFFFFF}{\color[HTML]{FF0000} \textbf{B2401}}} & \multicolumn{1}{c|}{\cellcolor[HTML]{FFFFFF}{\color[HTML]{FF0000} \textbf{New York}}} & \multicolumn{1}{c|}{{\color[HTML]{FF0000} \textbf{Wednesday}}} & \multicolumn{1}{c|}{{\color[HTML]{FF0000} \textbf{15:30}}} &            \\ \hline
\multicolumn{1}{|c|}{\cellcolor[HTML]{FFFFFF}E0823}                                 & \multicolumn{1}{c|}{\cellcolor[HTML]{FFFFFF}Tokyo}                                    & \multicolumn{1}{c|}{Friday}                                    & \multicolumn{1}{c|}{22:15}                                 & 1          \\ \hline
\multicolumn{1}{|c|}{\cellcolor[HTML]{FFFFFF}{\color[HTML]{FF0000} \textbf{Z1304}}} & \multicolumn{1}{c|}{\cellcolor[HTML]{FFFFFF}{\color[HTML]{FF0000} \textbf{Paris}}}    & \multicolumn{1}{c|}{{\color[HTML]{FF0000} \textbf{Monday}}}    & \multicolumn{1}{c|}{{\color[HTML]{FF0000} \textbf{9:45}}}  &            \\ \hline
\multicolumn{1}{|c|}{\cellcolor[HTML]{FFFFFF}L0620}                                 & \multicolumn{1}{c|}{\cellcolor[HTML]{FFFFFF}Sydney}                                   & \multicolumn{1}{c|}{Sunday}                                    & \multicolumn{1}{c|}{11:55}                                 & 1          \\ \hline
\multicolumn{1}{|c|}{\cellcolor[HTML]{FFFFFF}B1406}                                 & \multicolumn{1}{c|}{\cellcolor[HTML]{FFFFFF}Rome}                                     & \multicolumn{1}{c|}{Thursday}                                  & \multicolumn{1}{c|}{13:10}                                 & 2          \\ \hline
\multicolumn{5}{|c|}{\cellcolor[HTML]{FFFFFF}$\cdots \cdots$}                                                                                                                                                                                                                                                                       \\ \hline
\end{tabular}
}}

\end{table}

\section{Method}
\label{sec:method}
\begin{figure}[b]
    \centering
    \includegraphics[width=\linewidth]{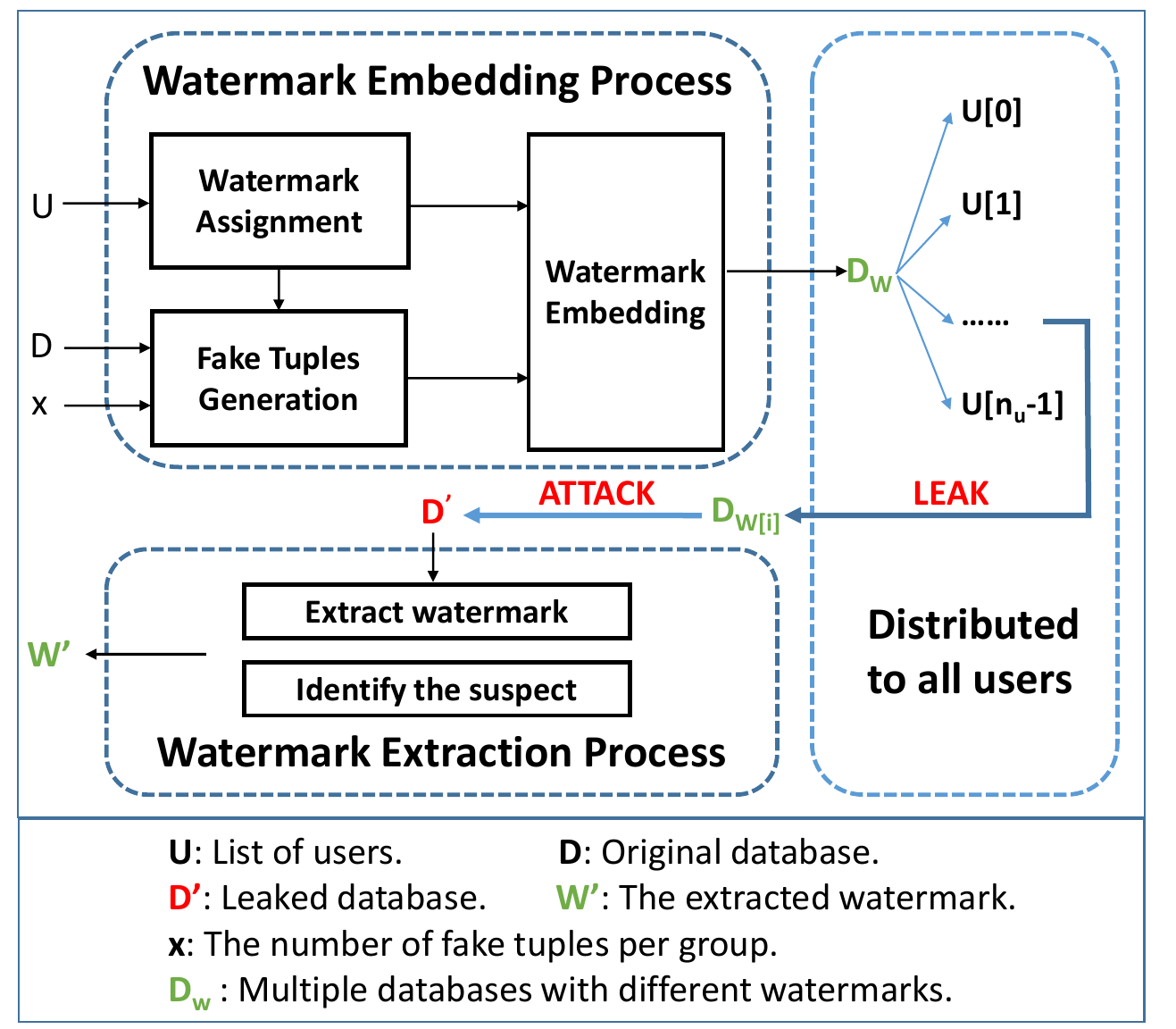}
    \caption{The framework of the proposed SPSW scheme.}
    \label{fig:kuang}
\end{figure}

In this section, we illustrate the details of the proposed SPSW, including the watermark embedding process and the extraction process. The framework is shown in Fig.~\ref{fig:kuang}.

\subsection{Watermark Embedding Process}
\subsubsection{Watermark Assignment}
In this paper, we use the binary sequence to represent watermark information. Assuming that there are $n_u$ database users, to trace all potential leakers, the length $L$ of the watermark should be $\lceil log_{2}(n_u) \rceil$. Typically, $n_u$ is less than $2^L$, so in practical applications, only a subset of the watermark sequence will be used. The proposed sparse priority strategy aims to prioritize the embedding of the sparsest watermark sequence from the pool of available sequences for each database user. Specifically, the watermark sequence with more `0' bits is regarded as sparser in this paper. 
For example, assuming there are three users, the watermark length should be set to $\lceil log_{2}(3) \rceil=2$. There are 4 watermark sequences with a length of 2, from the most sparse to the least sparse are ``00'', ``01'', ``10'' and ``11'', we first select the first three watermarks to distribute to these three users. The performance advantages brought by this strategy will be discussed in Section~\ref{subsec:EP}. Additionally, we will store mapping $M$ between users and watermark sequences, which has the following form:
\begin{equation}
    M = \{(U[i],\ W[i])\quad |\quad 0 \leq i \leq n_u - 1\},
\end{equation}
where $U[i]$ represents the $i$-th user, and $W[i]$ represents the watermark corresponding to the user.

%The detailed steps of the above process can be found in Algorithm \ref{alg:assi}.

% \begin{algorithm}[t]
%     \caption{Assign watermark}
%     \SetSideCommentRight
%     \label{alg:assi}
%     \KwIn{$U$(List of user names)}
%     \KwOut{$L$(Watermark length)\\
%            \qquad\quad\quad$T_c$(Correspondence table  between users and watermarks)}
%     $n_u \leftarrow \mathrm{len}(U)$\tcp{$n_u$ represents the number of users} 
%     $L \leftarrow \lceil \log_2(n_{u}) \rceil$ \tcp{Calculate the watermark length $L$}
%     $BinString \leftarrow \mathrm{ConstructAllBin}(0,2^{L})$ \tcp{Save all binary strings of length $L$ into $BinString$}
%     $SortedString \leftarrow \mathrm{sort}(BinString,key = BinString[i].\mathrm{count('1')})$ 
%     \tcp{Sort according to the number of 1s contained in each string from small to large}
%     \For{$i\leftarrow 0$ \KwTo $n_{u}-1$}{
%     $T_{c}[i][0] \leftarrow U[i]$ \\
%     $W_i \leftarrow SortedString[i]$ \\
%     $T_{c}[i][1] \leftarrow W_i$ \tcp{Correspond $n_u$ user name with the previous $n_u$ binary string}
%     }
% \end{algorithm}

\begin{algorithm}[t]
    \caption{Watermark Embedding }
    \SetSideCommentRight
    \label{alg:embed}
    \KwIn{\quad$L$ (Watermark length)\\
    \qquad\quad\quad$T_f$ (Generated fake tuples)\\
    \qquad\quad\quad$M$ (Mapping between user and watermark)\\
    \qquad\quad\quad$D$ (Original database)}
    \KwOut{$D_W$ (Multiple databases with different watermarks)}
    \For{$i\leftarrow 0$ \KwTo $n_u-1$}{
    $D_{W[i]} \leftarrow D$ 
    \tcp{$D_{W[i]}$ is the database of hidden watermark $W[i]$, which will be sent to the i-th user $U[i]$.}
    $W[i] \leftarrow M[i][1]$\\
    \For{$j\leftarrow 0$ \KwTo $L-1$}{
    \tcp*[h]{If the $j$-th watermark is '1', insert the fake tuples of the $j$-th group into the database} \\
    \If{$W[i][j] = \mathrm{'1'}$}{
    insert $T_f[j]$ to $D_{W[i]}$
    }
    % $D_W[i][0] \leftarrow T_{c}[i][0]$\\
    % $D_W[i][1] \leftarrow D_{W_i}$ \tcp{Record the correspondence between the watermarked database and the user}
    }
    }
\end{algorithm}

\subsubsection{Fake Tuples Generation}
To embed an $L$-bit watermark sequence using fake tuples, we generate $L$ groups of fake tuples for the database to be embedded. Each group corresponds to one bit in the watermark sequence and consists of $x$ fake tuples. Without loss of generality, natural language processing (NLP) models \cite{open2023} can assist us in completing this task. Specifically, by providing all the attribute information $A$ of the database $D$, the NLP model can imitate the real tuple data $t_r$ and generate multiple rows of fake data $d_f$. Then, by formatting the fake data, we can construct fake tuples and group them based on the parameters $L$ and $x$ to obtain fake tuples $T_f$, whose structure is as follows:

\subsubsection{Watermark Embedding}
Finally, for a specific database user $U[i]$, we query the correspondence table $M$ to obtain the watermark sequence $W[i]$ to be embedded. Following the proposed sparse priority embedding strategy, we will selectively embed the fake tuples corresponding to the `1' bits of the watermark sequence $W[i]$ into the database to obtain the watermarked database $D_{W[i]}$. Specifically, when the $j$-th bit of $W[i]$ is `1', all fake tuples in the group $T_f[j]$ will be embedded into the database. The illustration details of the above process are shown in Algorithm \ref{alg:embed}.

As shown in Table II, when using SPSW to embed the watermark $W=(10)_2$, it is only necessary to embed the first group of fake tuples in $T_f$ into the database. 
When $x=1$, only one fake tuple needs to be inserted. Compared with Table I, it can be observed that the proposed SPSW embeds fewer fake tuples under the same conditions, contributing to better transparency. In the next section, we have derived that the proposed SPSW has a smaller number of inserted fake tuples, whose theoretical upper bound is $xL/2$, and its better robustness is theoretically proved.

\renewcommand{\arraystretch}{1.4}
\begin{table}[t]
\centering
\caption{Embedding using SPSW Scheme.}
\label{tab:spsw}
\resizebox{\linewidth}{!}{
\setlength{\tabcolsep}{.1in}{
\begin{tabular}{|cccc|}
\hline
\multicolumn{1}{|c|}{\cellcolor[HTML]{FFFFFF}\textbf{Filght   number}} & \multicolumn{1}{c|}{\textbf{Department}}                            & \multicolumn{1}{c|}{\textbf{Day}}                                     & \textbf{Time}                                  \\ \hline
\multicolumn{1}{|c|}{Z0702}                                   & \multicolumn{1}{c|}{London}                                & \multicolumn{1}{c|}{Saturday}                                & 18:20                                 \\ \hline
\multicolumn{1}{|c|}{E0823}                                   & \multicolumn{1}{c|}{Tokyo}                                 & \multicolumn{1}{c|}{Friday}                                  & 22:15                                 \\ \hline
\multicolumn{1}{|c|}{{\color[HTML]{FF0000} \textbf{L3815}}}   & \multicolumn{1}{c|}{{\color[HTML]{FF0000} \textbf{Dubai}}} & \multicolumn{1}{c|}{{\color[HTML]{FF0000} \textbf{Tuesday}}} & {\color[HTML]{FF0000} \textbf{19:45}} \\ \hline
\multicolumn{1}{|c|}{L0620}                                   & \multicolumn{1}{c|}{Sydney}                                & \multicolumn{1}{c|}{Sunday}                                  & 11:55                                 \\ \hline
\multicolumn{1}{|c|}{B1406}                                   & \multicolumn{1}{c|}{Rome}                                  & \multicolumn{1}{c|}{Thursday}                                & 13:10                                 \\ \hline
\multicolumn{4}{|c|}{$\cdots \cdots$}                                                                                                                                                                                                          \\ \hline
\end{tabular}
}}
\end{table}

\subsection{Watermark Extraction Process}%Forensic Stage}
When a database appears in an unauthorized channel, we will extract the watermark information $W^{'}$ from this potentially compromised database $D^{'}$. And determine the main suspect through the corresponding relationship saved in the mapping $M$. Specifically, based on the fake tuples $T_f$ of $L$ groups, we will sequentially check whether the $j$-th group of fake tuples exists in the database to determine whether the $j$-th bit of the extracted watermark $W'$ is `1' or `0'.
Based on the extracted watermark $W^{'}$ we can find the prime suspect $s_p$. The specific extraction process can refer to Algorithm \ref{alg:extract}. 

% \begin{algorithm}[t]
%     \caption{Watermark Extraction Process }
%     \SetSideCommentRight
%     \label{alg:extract}
%     \KwIn{\quad$D^{'}$ (Leaked and attacked database)\\
%     \qquad\quad\quad$T_f$ (Generated fake tuples)\\
%     \qquad\quad\quad$M$ (Mapping between user and watermark)}
%     \KwOut{$W^{'}$ (Extracted watermark)\\
%     \qquad\quad\quad$s_p$ (Prime suspect)}
    
%     \For{$j\leftarrow 0$ \KwTo $L-1$}{
%     $flag \leftarrow 0$\\
%     \For{$k\leftarrow 0$ \KwTo $x-1$}{
%     $FakeTuple \leftarrow T_{f}[j][k]$\\
%     \If{$FakeTuple$ in $D^'$}{
%     $W^{'}[j] \leftarrow 1$\\
%     $flag \leftarrow 1$\\
%     break\\
%     }
%     }
%     \If{$flag = 0$}{
%     $W^{'}[j] \leftarrow 0$
%     }
%     }
    
%     \For{$i\leftarrow 0$ \KwTo $n_{u}-1$}{
%     \If{$W^{'} = M[i][1]$}{
%     $s_p \leftarrow M[i][0]$
%     }
%     }
    
% \end{algorithm}

\section{Theoretical Analysis}
\label{sec:theore}
Researchers usually use metrics such as robustness, transparency, and space-time complexity to evaluate the performance of watermarking schemes. In this section, we conduct a quantitative analysis of our proposed SPSW on these aspects. 

\subsection{Transparency Analysis}
\label{subsec:trans}
Watermark transparency refers to the degree of impact on the original data after the watermark embedding. High transparency means that the watermark is imperceptible and barely affects daily usage.

As mentioned in Section~\ref{sec:intro}, embedding watermarks by inserting fake tuples will not affect the original data, but may still have some impact on the use of the database. Generally, inserting fewer fake tuples means better transparency, so we use the average number $NI$ of inserted fake tuples per database to evaluate the transparency of the scheme. For a watermark of length $L$, if all watermark sequences are used with equal probability, the expected number of `1' bits for any unknown watermark sequence will be $L/2$. Because the proposed sparse priority strategy prioritizes embedding watermark sequences with fewer `1' bits, the expected number of `1' bits in the SPSW watermark sequence is less than $L/2$. 
%there are on average $L/2$ 1s in each string. When we preferentially use strings with fewer 1s as watermarks, the average number of 1s contained in each watermark will be lower than $L/2$. 
This means that each database only needs to insert fewer than $L/2$ sets of fake tuples on average. So $NI$ satisfies:
\begin{equation}
\label{equ:ni}
    NI < x\times \frac{L}{2} = \frac{xL}{2} = \frac{x \lceil \mathrm{log}_{2}n_{u} \rceil}{2}.
\end{equation}
Among them, $x$ is the number of fake tuples in each group of fake tuples. In contrast, existing fake-tuple-based database watermarking schemes typically require embedding $xL$ fake tuples, indicating that the proposed scheme offers better transparency.

\subsection{Robustness Analysis} 
Robustness refers to the ability of the watermarking scheme to accurately extract the watermark after distortions, especially the malicious attacks by leakers. Among various attacks, deletion attack is the most effective way, which destroys the watermark by deleting part of the tuples in the watermarked database. The focus of this section is to analyze the robustness of the proposed scheme under deletion attacks.

Generally, attackers are unable to distinguish between fake tuples and original ones, which forces them to destroy the watermark by randomly deleting tuples. Therefore, in the following discussion, we will analyze the robustness of the proposed scheme against random deletion attacks.
Additionally, we assess the robustness of the watermarking schemes by evaluating the probability expectation $EP$ of accurately extracting the watermark. In the following analysis, a higher value of $EP$ indicates a higher accuracy in watermark extraction, indicating a stronger level of robustness to random deletion attacks.

\subsubsection{Probability that a certain bit is right}
There are $x$ fake tuples in each group of fake tuples, and there are $n$ tuples in the database after embedding. When the deletion ratio is $p$, we need to calculate the probability $P_{cd}$ of a certain group of fake tuples being completely deleted.

Firstly, after the deletion attack, there are $(1-p)n$ tuples remaining, and $P_{cd}$ is equal to the probability that none of the fake tuples from a given group is included in the remaining $(1-p)n$ tuples. Considering that the given group has $x$ fake tuples, $P_{cd}$ can be calculated as follows:
\begin{equation}
\label{equ:pcd}
    \begin{aligned}
        P_{cd}
        &={\binom{n-x}{(1-p)n}} \div {\binom n{(1-p)n}}\\[2mm]
        &=\;\frac{(n-x)!\times \lbrack(1-p)n\rbrack!\times\lbrack n-(1-p)n\rbrack!}{\lbrack(1-p)n\rbrack!\times\lbrack n-x-(1-p)n\rbrack!\times n!}\\[2mm]%\div \frac{n!}{\lbrack(1-p)n\rbrack!\times\lbrack n-(1-p)n\rbrack!}
        &=\frac{(n-x)!}{n!}\times\frac{\lbrack(1-p)n\rbrack!\times(pn)!}{\lbrack(1-p)n\rbrack!\times(pn-x)!}\\[2mm]
        &=\frac{pn\times(pn-1)\times\dots\times(pn-x+1)}{n\times(n-1)\times\dots\times(n-x+1)}\\[2mm]&=\frac{pn}n\times\frac{pn-1}{n-1}\times\dots\times\frac{pn-(x-1)}{n-(x-1)}\\[2mm]&\lesssim\frac{pn}n\times\frac{pn-p}{n-1}\times\dots\times\frac{pn-p(x-1)}{n-(x-1)}\\[2mm]&=p^x.
        % &=\frac{pn\times(pn-1)\times\dots\times (pn-x+1)}{n\times(n-1)\times\dots\times(n-x+1)}\\[2mm]&=\frac{pn}n\times\frac{pn-1}{n-1}\times\dots\times\frac{pn-x+1}{n-x+1}\\[2mm]
        % &=p\times\frac{p(n-1)-(1-p)}{n-1}\times\dots\times\frac{p(n-x+1)-(x-1)(1-p)}{n-x+1}\\[2mm]
        % &=p\times(p-\frac{1-p}{n-1})\times\dots\times(p-\frac{(x-1)(1-p)}{n-x+1})\\[2mm]
        % &\lesssim p^x.
    \end{aligned}
\end{equation}

The probability $P_1$ of the bit `1' being correctly extracted is equal to the probability that the corresponding group of fake tuples is not completely deleted, that is, $P_1 = 1 - P_{cd}$. 
According to \eqref{equ:pcd}, $P_{cd}$ is less than and approximately equal to $p^x$, so $P_1$ is greater than and approximately equal to $1-p^x$.

For a bit `0' in the watermark sequence, due to the sparse priority embedding strategy, it will not be mistakenly extracted even under common database attacks. Therefore, the probability of the bit `0' being correctly extracted $P_0$ is always 1.

In conclusion, the probability of extracting a watermark bit correctly satisfies the following equation:
\begin{equation}
\label{equ:p1}
    P_1 = 1 - P_{cd} \gtrsim 1 - p^x,
\end{equation}
\begin{equation}
    \label{equ:p2}
    P_0 = 1.
\end{equation}

\subsubsection{Probability that the extracted watermark is right}
If we want to accurately extract a watermark $W$, each bit of the watermark must be extracted correctly. The probability of accurately extracting a watermark sequence with $k$ `1' bits can be denoted as $P_{ka}$ and is defined as follows: 
%Assuming that there are $k$ `1's and $(L-k)$ `0's in the watermark sequence, the probability $P_{ka}$ of accurate watermark extraction can be expressed as:
\begin{equation}
    \label{equ:pka}
    P_{ka} = P_1^k \times P_0^{(L-k)} = (1 - P_{cd})^k \times 1^{(L-k)} \gtrsim (1-p^x)^k.
\end{equation}

\subsubsection{Probability that there are $k$ `1's in one watermark}
Among the total $2^L$ watermarks of length $L$, there are $\binom{L}{k}$ watermarks that contain exactly $k$ `1' bits. Thus, the probability $P_{ko}$ of one watermark sequence having $k$ `1' bits can be given by:
%There are a total of $2^L$ watermarks of length $L$, among which $\binom{L}{k}$ watermarks contain $k$ 1s. Therefore, the probability $P_{ko}$ of a watermark with $k$ 1s occurring is:
\begin{equation}
    P_{ko} = \binom{L}{k} / 2^L.
\end{equation}

\subsubsection{Probability Expectation of accurately extracting the watermark}
\label{subsec:EP}
According to the above derivation, the probability expectation $EP$ of accurately extracting the watermark is:
\begin{equation}
    \label{equ:ep}
    \begin{aligned}
        EP &= \sum_{k=0}^{k=L}P_{ka}\times P_{ko} \\[2mm]
        &= \sum_{k=0}^{k=L}(1 - P_{cd})^k \times \frac{\binom{L}{k}}{2^L}\\[2mm]
        &=  \frac{1}{2^L} \times \sum_{k=0}^{k=L}\binom{L}{k} \times (1 - P_{cd})^k \times 1^{(L-k)}\\[2mm]
        &= \frac{1}{2^L}  \times (1 - P_{cd} + 1)^L \\[2mm]
        &= (1 - \frac{1}{2}P_{cd})^L.
    \end{aligned}
\end{equation}
As mentioned in Section~\ref{sec:intro}, the number of users that need to be traced is usually smaller than the information capacity of the watermark, which is $2^L$. According to the proposed sparse priority embedding strategy, our proposed watermark scheme will prioritize embedding watermark sequences with more `0' bits. According to \eqref{equ:pka}, the sparse priority embedding strategy leads to a smaller $k$, which in turn increases the probability expectation $EP$ of accurately extracting the watermarks. Thus, in most cases, the probability expectation $EP$ can be approximated as follows:
\begin{equation}
\label{eq:EP}
    EP > (1 - \frac{1}{2}P_{cd})^L.
\end{equation}

\subsubsection{Robustness comparison with existing methods}
Existing fake-tuple-based database watermarking schemes typically employ more traditional methods to embed the watermark sequence. Specifically, for each bit in the watermark sequence, a corresponding group of fake tuples is embedded into the database based on whether the bit is `0' or `1'. When facing deletion attacks, if a specific group of fake tuples is completely deleted, the corresponding watermark bit will be randomly set to 0 or 1, resulting in a potential 50\% chance of incorrect extraction.
Therefore, in existing schemes, the probability that a certain bit is wrong is equal to half the probability that the corresponding fake tuples group is completely deleted, which is $\frac{1}{2}P_{cd}$. Thus, we can conclude that the probability of a certain bit being extracted correctly in existing fake-tuple-based database watermarking schemes is:
\begin{equation}
    P_0^{'} = P_1^{'} = 1 - \frac{1}{2}P_{cd}.
\end{equation}
For a watermark with a length of $L$, it is necessary to achieve a completely correct extraction, that is, each bit is extracted correctly. For a watermark with $k$ `1's, the probability of correct watermark extraction $P_a^{'}$ is:
\begin{equation}
    P_a^{'} = (P_1^{'})^k \times (P_0^{'})^{L-k} = (1 - \frac{1}{2}P_{cd})^L.
\end{equation}
It can be observed that in existing watermarking methods, the probability of correct watermark extraction is independent of the distribution of `0/1' bits.
Therefore, the expectation $E_P^{'}$ for the watermark to be accurately extracted is:
\begin{equation}
\label{eq:EP'}
    EP^{'} = P_a^{'} \times 1 = (1 - \frac{1}{2}P_{cd})^L.
\end{equation}

Combining \eqref{eq:EP} and \eqref{eq:EP'}, it is obvious that $EP > EP^{'}$. Thus, compared with the existing fake-tuple-based database watermarking scheme, the proposed SPSW has a higher probability expectation of correct extraction, i.e., better robustness.

%compared with the traditional method based on modification, the probability of accurate watermark extraction in this scheme is expected to be higher. Therefore, this method has stronger robustness.

Additionally, it should be noted that the proposed sparse priority strategy also enhances the traceability of the proposed watermarking scheme. Specifically, when the watermarked database is subjected to attacks or distortions leading to incorrect watermark extraction, the proposed watermarking scheme can provide a smaller set of potential leak suspects. For example, when $L=4$ and the incorrectly extracted watermark sequence is 1010, we can assert that users corresponding to 1110 and 1011 are more likely to be potential leak suspects compared to the user corresponding to 1111. However, existing database watermarking schemes are unable to narrow down the set of suspects further, because their corresponding watermark sequences have the same correct probability.

\subsection{Complexity Analysis}%Space-Time Complexity}
Time complexity and space complexity are also important metrics for evaluating the practical application of database watermarking methods, as they assess the efficiency and resource consumption of the proposed scheme. In addition to the common time complexity, fake-tuple-based database watermarking methods typically require storing the fake tuples to facilitate watermark embedding and extraction. Therefore, the space complexity of such schemes also needs to be evaluated.
%We need to evaluate the efficiency and resource consumption of this scheme. The efficiency of an algorithm can be evaluated using the time complexity. The scheme needs to store some files to ensure the embedding and extraction of the watermark, which will involve a certain amount of storage space consumption, so the space complexity of the stored files needs to be used to evaluate the resource consumption.

\subsubsection{Time Complexity of Embedding Process}
As mentioned in Section~\ref{sec:method}, the proposed watermark embedding process can be divided into three steps, including watermark assignment, fake tuples generation, and watermark embedding. 
In the watermark assignment, the $2^L$ watermark sequences of length $L$ need to be sorted, and $n_u$ of them are selected to map to $n_u$ users. Thus, the time complexity of the assignment process is $O_a$:
\begin{equation}
    O_a = O(2^L\log_2(2^L) + n_u) = O(n_u\log{n_u}).
\end{equation}

Then, in the fake tuples generation, we fed the attributes and some real tuples of the database into the NLP model to generate imperceptible fake tuples. In this paper, the proposed SPSW requires $L$ groups of fake tuples and each group has $x$ fake tuples. The time complexity of this step is $O_g$:
\begin{equation}
    O_g=O(xL)=O(x\log{n_{u}}).
\end{equation}

%it is necessary to pass the attributes of the database and some real tuples as parameters to the NLP model, obtain the generated fake data, and organize them into $x$ fake tuples for each group, a total of $L$ groups. The time complexity of this link is O(generate) =O($xL$) =O($xlog{n_{u}}$).

Finally, in the watermark embedding step, for each of the $n_{u}$ users, iterate through their corresponding watermarks and decide whether to insert fake tuples based on the state of each watermark bit. The time complexity of this step is $O_e$:
\begin{equation}
    O_e = O(n_{u}L) =O(n_{u}\log{n_{u}}).
\end{equation}

Therefore, the time complexity of the embedding process is determined by the maximum value among the time complexities of the three aforementioned steps.
%Therefore, the maximum value of the time complexity of the above three steps is regarded as the time complexity of the embedding stage.
\begin{equation}
    \begin{aligned}
         O(\textbf{Embedding}) %&= %max(O(\textbf{assigning}),O(\textbf{generate}),O(\textbf{embedding%}))\\
        &=O(n_{u}\log{n_{u}}).%\nonumber
    \end{aligned}
\end{equation}

\subsubsection{Time Complexity of Extraction Process}
In the watermark extraction process, the watermark bit is determined by detecting whether any fake tuple of a certain group is contained in the database. Then, based on the extracted watermark bits, identify the corresponding user. So the time complexity of the extraction process is:
\begin{equation}
    \begin{aligned}
        O(\textbf{Extraction})&=
        O(n_{u}xL)\\&=O(xn_{u}\log{n_{u}}).%\nonumber
    \end{aligned}
\end{equation}

\subsubsection{Space Complexity}
Like existing fake-tuple-based database watermarking schemes, the proposed SPSW also requires storing two tables. One is the fake tuple list, including $xL$ fake tuples. Another is the mapping table between watermarks and users, which stores $n_u$ users and their corresponding watermarks. So the space complexity of this scheme is:
\begin{equation}
    \begin{aligned}
        O(\textbf{Space complexity}) &=
        O(xL + n_{u})\\&=O(x\log{n_{u}}+n_{u}).%\nonumber
    \end{aligned}
\end{equation}

The space-time complexity of the traditional method is related to the number of tuples $n$, but the space-time complexity of SPSW is related to the number of users $n_u$, which is much lower than the traditional method. Therefore, this scheme has obvious advantages in terms of time and space complexity.

\section{Experiment}
\label{sec:experiment}
In this section, we verify the robustness and transparency of this scheme through experiments and compare it with schemes PEEW \cite{farfoura2013novel}, GAHSW \cite{hu2018new} and FBIPT \cite{wang2022fbipt}.

In the experiment, we used the open-source Traffic Violations in Maryland County dataset \cite{dataset}, added a primary key to each piece of data, and saved it as a database as the carrier of the watermarking scheme. The database contains about 1.04 million tuples with attributes including accident time, accident description, accident location, longitude, latitude, etc.

In order to ensure that the experiment is performed under the same conditions, we set the default values of the experimental parameters: The number of tuples in the database $n=10000$, the number of database users $n_u=50$, and the number of fake tuples in each set of fake tuples $x=5$. 
%$n$=10000, $n_u$=50, $x$=5. 
This allows different schemes to be compared under the same conditions to evaluate their performance.

\subsection{Robustness Experiment}

\begin{figure}[t]
    \centering
    \includegraphics[width=0.5\textwidth,height = 0.4\textwidth]{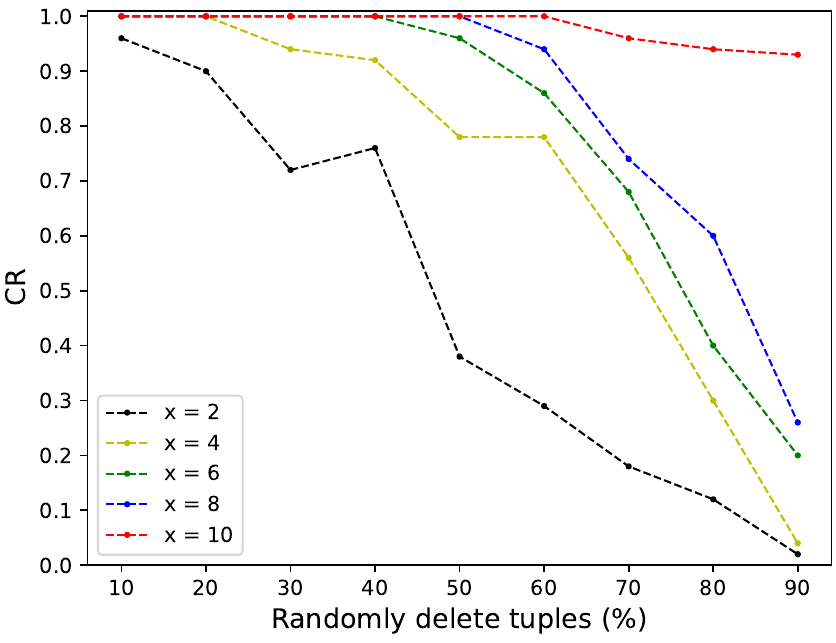}
    \caption{The effect of the number $x$ of fake tuples per group on the extraction accuracy.}
    \label{fig:x}
\end{figure}
According to \eqref{equ:ep}, the probability expectation that the watermark is correctly extracted is affected by the erasure attack ratio $p$, the watermark length $L$, and the number $x$ of fake tuples per group. 
The watermark length $L$ depends on the number of users $n_u$, so we only discuss $n_u$ later on. To verify the effect of these parameters on the probability expectation of correct extraction, we conduct experimental observations and judge whether the experimental results are consistent with the theoretical derivation.

In the first set of experiments, we fix the number of users $n_u$ and embed watermark with different parameters $x$. For each watermarked database, we randomly delete tuples of $p$ proportion, which ranges from 10\% to 90\%. 
To ensure the reliability of the experimental results, we repeated 50 trials for each combination of $x$ and $p$ and calculate the correct rate $CR$ of each combination. The experimental results are shown in Fig.~\ref{fig:x}.

\begin{figure}[t]
    \centering
    \includegraphics[width=0.5\textwidth,height = 0.4\textwidth]{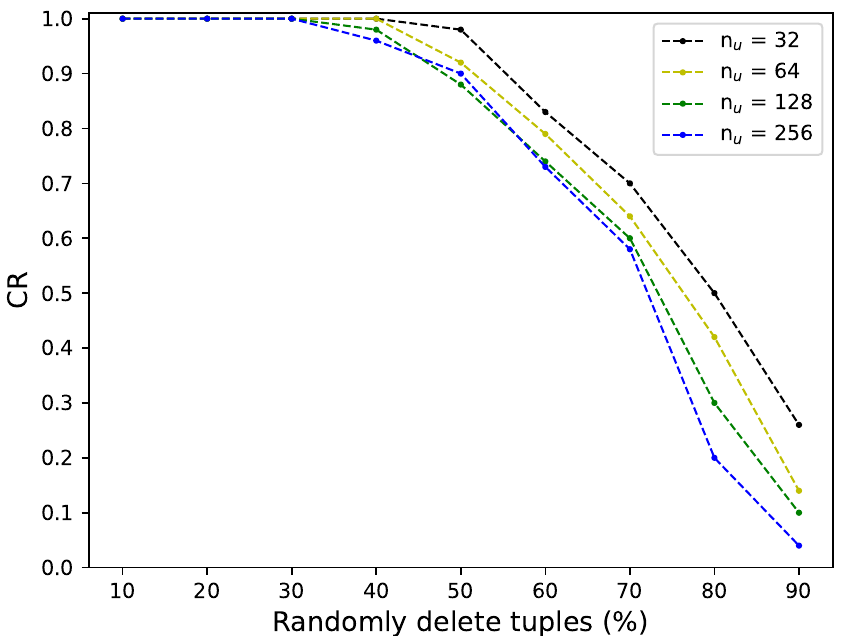}
    \caption{The effect of the number of users $n_u$ on the extraction accuracy.}
    \label{fig:nu}
\end{figure}

%In Fig.~\ref{fig:x}, the horizontal axis represents the deletion ratio $p$, and the vertical axis represents the correct extraction ratio $CR$, and different lines represent different parameters $x$. 
It can be observed that as the deletion ratio $p$ increases, the correct rate $CR$ presents a downward trend; and as the parameter $x$ increases, the correct rate $CR$ presents an upward trend. This is consistent with our previous theoretically derived results. When each group of fake tuples consists of 10 tuples, corresponding to embedding 0.3\% fake tuples in the default setting, $CR$ is still very close to 1 even if 60\% of the database tuples are deleted. When 90\% of the tuples are deleted, $CR$ is still higher than 0.9. This shows that the scheme can maintain high robustness in the face of serious deletion attacks.

Then, we evaluate the effect of the number of users $n_u$ on the watermark extraction process. The experimental conditions were the same as the previous set of experiments, with $n_u$ being used as the independent variable. 
%In the second set of experiments, we will investigate the effect of the number of users on the correct rate. The experiment flow is the same as before, the only changes are fixing the number of fake tuples and changing the number of users. 
The experimental results are shown in Fig.~\ref{fig:nu}. 
%Different lines in the figure represent different numbers of users. 
It can be observed that as the number of users $n_u$ increases, the correct rate $CR$ presents a downward trend, which is consistent with our theoretical derivation results.

\begin{figure}[t]
    \centering
    \includegraphics[width=0.5\textwidth,height = 0.4\textwidth]{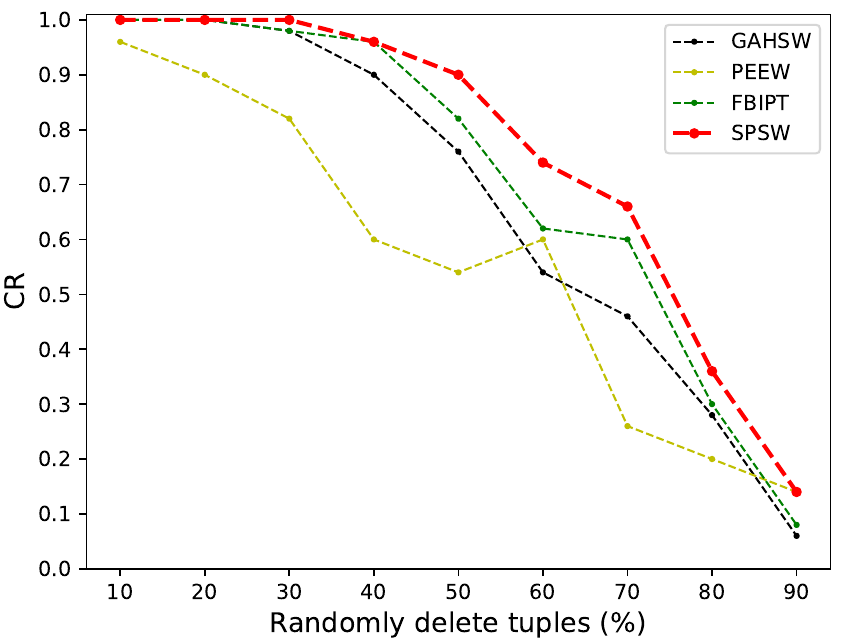}
    \caption{Comparison of the extraction accuracy between comparative schemes and the proposed SPSW.}
    \label{fig:ran}
\end{figure}

Finally, a comparative experiment is conducted. Under the same experimental settings, we randomly delete tuples from watermarked databases generated by both the comparative watermarking scheme and the proposed scheme and extract the watermark. Fig.~\ref{fig:ran} shows the average extraction performance of 50 repetitive experiments. 
%Under the same conditions, different schemes are used for watermark embedding, and different proportions of tuples are randomly deleted for watermark extraction. The experiment will be repeated 50 times and the proportion of correct extractions will be calculated. The experimental results are shown in Fig.~\ref{fig:ran}.

\subsection{Transparency Experiment}
\begin{figure}[t]
\centering
\includegraphics[width=0.5\textwidth,height = 0.4\textwidth]{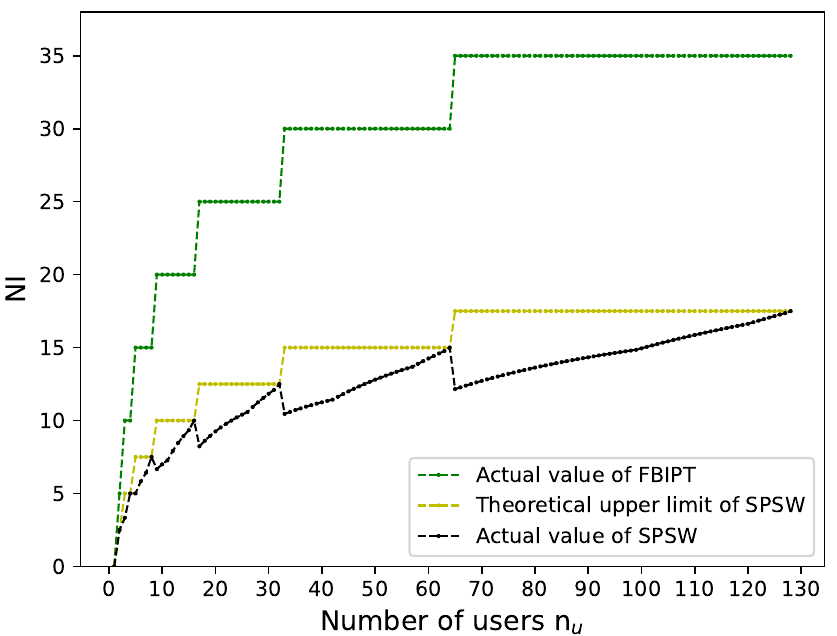}
\caption{Average value of inserted fake tuples.} \label{fig:touming}
\end{figure}

We evaluate the transparency of one database watermarking scheme by the average number $NI$ of inserted fake tuples in the database. According to the analysis in Section\ref{subsec:trans} and \eqref{equ:ni}, the actual $NI$ value would be lower than the theoretical upper limit. During the experiments, we performed the embedding algorithm for different numbers of users and recorded the corresponding $NI$ values. The experimental results are shown in Fig.~\ref{fig:touming}.

In Fig.~\ref{fig:touming}, the horizontal axis represents the number of users, and the vertical axis represents the average number of inserted fake tuples $NI$. The yellow line represents the theoretical upper limit, and the black line represents the actual value of $NI$. 
It can be observed that the black line is always below the yellow line, indicating that the number of tuples inserted is less than the theoretical upper limit, which aligns with the conclusion of theoretical derivation in \eqref{equ:ni}. At the same time, the theoretical upper limit of our proposed SPSW is half of the $NI$ of existing schemes, which demonstrates the superiority of our scheme in terms of transparency.

\section{Conclusion}
\label{sec:conclusion}
This paper proposes a traceable database watermarking scheme based on the sparsity priority strategy, named SPSW, which realizes the traceability function by inserting fake tuples. This scheme uses the NLP model to generate fake tuples similar to real data and uses a sparse priority strategy to improve robustness and transparency. Theoretical analysis and experiments show that the scheme has strong robustness and accurate watermark extraction ability, and provides a feasible and effective solution to the traceability problem. Future research can further optimize the algorithm, explore more attack scenarios, and combine other technical means to improve the security and reliability of the scheme.
\printbibliography
\end{document}